\documentclass[12pt, a4paper]{article}
\usepackage{a4wide,amssymb,amsmath,amscd,}

\newcommand{\T}{{\mathbb T}}
\newcommand{\Z}{{\mathbb Z}}
\newcommand{\R}{{\mathbb R}}
\newcommand{\C}{{\mathbb C}}

\begin{document}

\topmargin -2pt

\headheight 0pt

\topskip 0mm \addtolength{\baselineskip}{0.20\baselineskip}
\begin{flushright}
{\tt math-ph/0605075} \\
{\tt KIAS-P06005} \\
\end{flushright}


\begin{center}
{\Large \bf Quantum Thetas on Noncommutative ${\T}^4$  \\
  from Embeddings into Lattice } \\

\vspace{10mm}

{\sc Ee Chang-Young}\footnote{cylee@sejong.ac.kr}\\
{\it Department of Physics, Sejong University, Seoul 143-747, Korea}\\

{\it School of Physics, Korea Institute for Advanced Study,
Seoul 130-722, Korea}\\

\vspace{2mm}

and \\

\vspace{2mm}

{\sc Hoil Kim}\footnote{hikim@knu.ac.kr}\\

{\it Topology and Geometry Research Center, Kyungpook National University,\\
Taegu 702-701, Korea}\\

\vspace{10mm}

{\bf ABSTRACT} \\
\end{center}

\vspace{2mm}

\noindent In this paper we investigate the theta vector and
quantum theta function over noncommutative ${\T}^4$ from the
embedding of $ {\R} \times {\Z}^2 $. Manin has constructed the quantum
theta functions from the lattice embedding into vector space
($\times$ finite group). We extend Manin's construction of the quantum
theta function to the embedding of vector space $\times$ lattice
case. We find that the holomorphic theta vector exists only over
the vector space part of the embedding, and over the lattice part
we can only impose the condition for Schwartz function. The
quantum theta function built on this partial theta vector
satisfies the requirement of the quantum theta function. However, two
subsequent quantum translations from the embedding into the
lattice part are non-additive, contrary to the additivity of those
from the vector space part.
\\


\noindent

\thispagestyle{empty}

\newpage
\section*{1. Introduction}

In the quantization of a classical theta function, we encounter two
types of objects. One is the theta vector introduced by
Schwarz\cite{schwarz01}, which is a holomorphic element of a
projective module over a unitary quantum torus. The other is the
quantum theta function introduced by
Manin\cite{manin1,manin1-tr,manin2,manin3}, which is an element of
the function ring of the quantum torus itself. This is a natural
outcome if one considers the process of quantization, in which
commutative physical observables become operators acting on the
states. Classically one deals with one type of objects,
observables. After quantization one deals with two types of
objects, operators and states. That is what happens here. In the
classical sense, a set of specific values of observables
constitutes a state. The (classical) theta function is just like a
state function. On the other hand, the quantum theta function and
theta vector correspond to an operator and a state vector,
respectively, in the quantum sense. Manin\cite{manin2,manin3} has
defined the quantum theta function via the Rieffel's algebra
valued inner product \cite{rief88} of a theta vector \cite{ds02}
from the embedding of the type $ {\R}^p (\times F)$ for the quantum
torus.  Here, $d= 2p $ is the dimension of the relevant quantum
torus and $F$ is a finite group. In \cite{rief88} it was shown
that the general embedding for the quantum torus is of the type $
{\R}^p \times {\Z}^q (\times F )$, where $d= 2p + q$ is the
dimension of the relevant quantum torus. Manin has constructed the
quantum theta functions only for the embeddings of  $ {\R}^p $
type, and those from the $ {\R}^p \times {\Z}^q$ type have been
left in question \cite{manin3}.

One needs to know the result of the ${\Z}^q$ type embedding in
order to understand the full symmetry of quantum tori including
the Morita equivalence. In \cite{ek05}, the symmetry of the quantum
torus was investigated, restricted to the symmetry of the algebra
and its module, not related to the Morita equivalence.
In this paper, we construct the quantum theta functions in a more
general $ {\R}^p \times {\Z}^q$  type of embedding that Manin did
not investigate. We first investigate the existence of the theta
vector in this setup, and find that the holomorphic theta vector
does not exist in the exact sense. It turns out that one can only
construct partially holomorphic theta vectors, which are
holomorphic for the embedding into the vector space ($ {\R}^p $)
part but not for the lattice (${\Z}^q$) part.
We then investigate whether the quantum theta function satisfying
the Manin's requirement can be constructed with this partially
holomorphic theta vector. We find that the answer is yes.
The organization of this paper is as follows. In section 2, we
construct a module for the quantum 4-torus with the embedding of $
{\R}^p \times {\Z}^q$ type. In section 3, we construct the
quantum theta function evaluating the scalar product of the above
 module, and check the Manin's requirement for the quantum
theta function. In section 4, we conclude with the discussion.




\section*{2. Lattice embedding of the quantum torus}\label{embedd}

Here we first review the embedding of the quantum torus
\cite{rief88} and an explicit construction of the module with an
embedding of the type  $ {\R}^p (\times F )$ which was done for
the 4-torus case in \cite{kl03}. Then we construct the module
with an embedding of the type  $ {\R}^p \times {\Z}^q (\times F )$
for the quantum 4-torus.

Quantum torus ${\T}^d_\theta$ is a deformed algebra of the algebra
of smooth functions on the torus ${\T}^d$ with the deformation
parameter $\theta$, which is a real $d\times d$ anti-symmetric
matrix. This algebra is generated by operators $U_1,\cdots,U_d$
obeying the following relations
\begin{align*}
U_jU_i=e^{2\pi i \theta_{ij}}U_iU_j \text{ \ and \ }
U_i^*U_i=U_iU_i^*=1, \text{ \ \ } i,j=1,\cdots,d.
\end{align*}
The above relations define the representation of the involutive
algebra
$${\cal A}_\theta^d=
\{\sum a_{i_1\cdots i_d}U_1^{i_1}\cdots U_d^{i_d}\mid
a=(a_{i_1\cdots i_d})\in {\cal S}({\Z}^d)\}$$ where ${\cal
S}({\Z}^d)$ is the Schwartz space of sequences with rapid decay.

Every projective module over a smooth algebra ${\cal
A}^{d}_{\theta}$ can be represented by a direct sum of modules of
the form ${\cal S}({\R}^p\times{\Z}^q\times F)$, the linear space
of Schwartz functions on ${\R}^p\times{\Z}^q\times F$, where
$2p+q=d$ and $F$ is a finite abelian group.
 The module action is specified
by operators on ${\cal S}({\R}^p\times{\Z}^q\times F)$ and the
commutation relation of these operators should be matched with
that of elements in ${\cal A}^{d}_{\theta}$.

 Recall that there is the dual action of
the torus group ${\T}^d$ on ${\cal A}_\theta^d$ which gives a Lie
group homomorphism of ${\T}^d$ into the group of automorphisms of
${\cal A}_\theta^d$. Its infinitesimal form generates a
homomorphism of Lie algebra $ L$ of ${\T}^d$ into Lie algebra of
derivations of ${\cal A}_\theta^d$. Note that the Lie algebra $L$
is abelian and is isomorphic to ${\R}^d$. Let $\delta:L\rightarrow
{\rm {Der \ }}({\cal A}_\theta^d)$ be the homomorphism. For each
$X\in L$, $\delta(X):=\delta_X$ is a derivation i.e., for $u,v\in
{\cal A}_\theta^d$,
\begin{equation} \label{derv1}
\delta_X(uv)=\delta_X(u)v+u\delta_X(v).
\end{equation}
Derivations corresponding to the generators $\{e_1,\cdots,e_d\}$
of $L$ will be denoted by $\delta_1,\cdots,\delta_d$. For the
generators $U_i$'s of ${\T}_\theta^d$, it has the following
property
\begin{equation} \label{derv2}
\delta_i(U_j)=2\pi i\delta_{ij} U_j.
\end{equation}
Let $D$ be a lattice in $\cal{G}=M\times \widehat{M}$, where
$M={\R}^p\times{\Z}^q\times F$ and $\widehat{M}$ is its dual. Let
$\Phi$ be an embedding map such that $D$ is the image of ${\Z}^d$
under the map $\Phi$. This determines a projective module to be
denoted by $E$ \cite{rief88}. If $E$ is a projective ${\cal
A}_\theta^d$-module, a connection $\nabla$ on $E$ is a linear map
from $E$ to $E\otimes L^*$ such that for all $X\in L$,
\begin{align} \label{conn1}
\nabla_X(\xi u)=(\nabla_X\xi)u+\xi\delta_X(u),{\rm { \ \ \
}}\xi\in {E}, u\in {\cal A}_\theta^d.
\end{align}
It is easy to see that
\begin{align} \label{conn2}
[\nabla_i,U_j]=2\pi i\delta_{ij} U_j.
\end{align}

\noindent
In the Heisenberg representation the operators are
 defined by
\begin{equation}\label{pirep}
{\cal U}_{(m,\hat s)}f(r)=e^{2\pi i <r, \hat s>}f(r+m)
\end{equation}
for $(m,\hat s)\in\ D, \ r \in M .$

\subsection*{2.1 Embedding into vector space } \label{embqzero}

We now review the explicit construction of a module over
noncommutative ${\T}^4$ with the embedding of the type $ {\R}^2
(\times F)$ \cite{kl03}.

\noindent
 For the real part, we choose our
embedding map as
\begin{align}\label{phii}
\Phi_{\rm inf}=\begin{pmatrix}\theta_1 + \frac{n_1}{m_1}&0 &0&0\\
                0&0&\theta_2 + \frac{n_2}{m_2}&0\\
                0&1&0&0\\
                0&0&0&1\end{pmatrix} \equiv (x_{ij}).
\end{align}
Then using the previous expression for the Heisenberg
representation,
\begin{align}
(V_if)(s_1,s_2)=(V_{e_i}f)(s_1,s_2):=\exp(2\pi
i(s_1x_{3i}+s_2x_{4i}))f(s_1+x_{1i},s_2+x_{2i}), ~ s_1,s_2 \in \R , \notag
\end{align}
we get
\begin{align}
(V_1f)(s_1,s_2)&=f(s_1 + \theta_1 + \frac{n_1}{m_1}, s_2), \nonumber\\
(V_2f)(s_1,s_2)&=\exp( 2\pi i s_1 )f(s_1, s_2), \nonumber\\
(V_3f)(s_1,s_2)&=f(s_1,s_2 + \theta_2 + \frac{n_2}{m_2} ), \notag\\
(V_4f)(s_1,s_2)&=\exp(2 \pi i s_2)f(s_1,s_2).\nonumber
\end{align}

For the finite part, let $F={\Z}_{m_1}\times{\Z}_{m_2}$, where
${\Z}_{m_i}={\Z}/m_i{\Z}$, ($i=1,2$) and consider the space
${\C}^{m_1}\otimes{\C}^{m_2}$ as the space of functions on
$C({\Z}_{m_1}\times{\Z}_{m_2})$. For all $m_i\in {\Z}$ and $n_i\in
{\Z}/m_i{\Z}$ such that $m_i$ and $n_i$ are relatively prime,
we define the operators $W_i$ on $C({\Z}_{m_1}\times{\Z}_{m_2})$
corresponding to our embedding map
\begin{align}\label{phif}
\Phi_{\rm fin}=\begin{pmatrix} -1&0&0&0\\
                0&0&-1&0\\
                0& \frac{n_1}{m_1}&0&0\\
                0&0&0& \frac{n_2}{m_2}\end{pmatrix}
\end{align}
with $ k_i \in \Z_{m_i} \ (i=1,2) $ as follows:
\begin{align}
(W_1f)(k_1,k_2)&=f(k_1-1,k_2), \nonumber\\
(W_2f)(k_1,k_2)&=\exp(2\pi i \frac{n_1 k_1}{m_1})f(k_1,k_2), \nonumber\\
(W_3f)(k_1,k_2)&=f(k_1,k_2-1), \notag\\
(W_4f)(k_1,k_2)&=\exp(2 \pi i
\frac{n_2k_2}{m_2})f(k_1,k_2).\nonumber
\end{align}

\noindent
Now, we define operators $U_i=V_i\otimes W_i$ acting on
the space $E :={\mathcal S}({\mathbb R}^2)\otimes {\mathbb
C}^{m_1}\otimes{\C}^{m_2}$ as
\begin{align} \label{uopr}
(U_1f)(s_1,s_2,k_1,k_2)&=
                 f(s_1  + \theta_1 + \frac{n_1}{m_1},s_2,k_1-1,k_2), \notag \\
(U_2f)(s_1,s_2,k_1,k_2)&=e^{2\pi i(s_1 + \frac{n_1 k_1}{m_1} )}
                 f(s_1,s_2,k_1,k_2), \notag \\
(U_3f)(s_1,s_2,k_1,k_2)&=
                 f(s_1,s_2 + \theta_2 + \frac{n_2}{m_2},k_1,k_2-1), \notag \\
(U_4f)(s_1,s_2,k_1,k_2)&=e^{2\pi i(s_2 + \frac{n_2 k_2}{m_2} )}
                 f(s_1,s_2,k_1,k_2) .
\end{align}
One can see that they satisfy
\begin{align} \label{Ucomm}
U_2U_1 &=e^{2\pi i \theta_1}U_1U_2, \notag \\
U_4U_3 &=e^{2\pi i \theta_2}U_3U_4,
\end{align}
and otherwise $U_i U_j = U_j U_i $.


\subsection*{2.2 Embedding into lattice } \label{embqnzero}

Here, we do a similar construction for the embedding of the type $
{\R}^p \times {\Z}^q (\times F )$. The embedding of the finite
part can be done in the exactly same manner as in the previous
subsection. Thus we will suppress the expression for the finite
part for brevity, and only consider the infinite part with the
embedding of the type $ {\R}^p \times {\Z}^q $ with $p=1$ and
$q=2$.

\noindent Here, we embed $D \subset {\R}^4$ into $ {\R} \times
{\Z}^2 \times {\R}^* \times {\T}^2 $, and we choose our embedding
as follows:
\begin{align}\label{embqnz}
\Phi_{\rm inf}=\begin{pmatrix}\theta_1&0 &0&0\\
                0&0&m_{11}&m_{12}\\ 0&0&m_{21}&m_{22}\\ 0&1& 0& 0 \\
                0&0& \hat{\delta}_{11} & \hat{\delta}_{12}\\
                0&0&\hat{\delta}_{21} &\hat{\delta}_{22} \end{pmatrix} \equiv (x_{ij}),
\end{align}
where $\theta_1 \in {\R}$, and $m_{nl} \in {\Z} $, $
\hat{\delta}_{nl} \in {\T} $ for $n,l=1,2$, and $i=1,..,6,
~j=1,..,4$.
Then, the operators $U_j$ acting on the space $E :={\mathcal
S}({\mathbb R} \otimes {\Z}^2 ) $ can be defined as
\begin{align}
(U_1f)(s,n_1,n_2)&= f(s  + \theta_1, n_1,n_2), \notag \\
(U_2f)(s,n_1,n_2)&=e^{2\pi i s }
                 f(s,n_1,n_2), \notag \\
(U_3f)(s,n_1,n_2)&= e^{2\pi i (\hat{\delta}_{11} n_1 +
\hat{\delta}_{21} n_2 ) + \pi i (m_{11} \hat{\delta}_{11} + m_{21}
\hat{\delta}_{21} ) }
                 f(s,n_1 +m_{11},n_2+m_{21}), \notag \\
(U_4f)(s,n_1,n_2)&= e^{2\pi i (\hat{\delta}_{12} n_1 +
\hat{\delta}_{22} n_2 ) + \pi i (m_{12} \hat{\delta}_{12} + m_{22}
\hat{\delta}_{22} )  }
                 f(s,n_1 +m_{12},n_2+m_{22}),
\end{align}
where $ s \in {\R}, ~ n_{l} \in {\Z}$  for $l=1,2$.
In the above definition of $U_i$ operators, an extra phase term is added
to conform with Manin's definition of quantum theta function \cite{manin3}.
 The above can be compactly written as
\begin{align}
\label{uopr} (U_j f)(s,n_1,n_2)&= e^{2\pi i ( s x_{4j} + n_1
x_{5j}+ n_2 x_{6j} ) + \pi i ( \sum_{k=1}^3 x_{kj} x_{(k+3) j}) }
                 f(s+x_{1j},n_1 +x_{2j},n_2+x_{3j})
\end{align}
for $j=1,..,4$.

The commutation relations among $U_i$'s are given by
\begin{align} \label{Ucomm}
U_2U_1 &=e^{2\pi i \theta_{12}}U_1U_2, \notag \\
U_4U_3 &=e^{2\pi i \theta_{34}}U_3U_4,
\end{align}
where $\theta_{12} = \theta_1, ~~ \theta_{34} = m_{11}
\hat{\delta}_{12} + m_{21} \hat{\delta}_{22} - m_{12}
\hat{\delta}_{11} - m_{22} \hat{\delta}_{21} $, ~  and otherwise
$U_i U_j = U_j U_i $.

%
\section*{3. Quantum thetas}\label{qtheta}

In this section, we first define connections with complex
structures for the two embedding cases in the previous section and
consider the theta vector in each case. Then, we define the
quantum theta function for each case following the Manin's
construction.

%
\subsection*{3.1 Theta vectors}\label{thevec}




In the previous section, connections on a projective ${\cal
A}_\theta^d$-module satisfies the condition (\ref{conn2}) and it
can be written as
\begin{align}
\label{concr}
 U_j \nabla_i U_j^{-1} &= \nabla_i
- 2 \pi i \delta_{ij} .
\end{align}
With this condition in mind, now we construct the theta vector for
each embedding case.
%



\subsubsection*{3.1.1 Embedding into vector space}
%

 For the embedding of the type $ {\R}^2
(\times F)$, the above relation is satisfied, if we set
\begin{align*}
(\nabla_i f)(s_1,s_2)&= - 2\pi iA_{i1}s_1f(s_1,s_2) - 2\pi
iA_{i2}s_2f(s_1,s_2) \\
& + A_{i3}\frac{\partial f(s_1,s_2)}{\partial s_1}  +
A_{i4}\frac{\partial f(s_1,s_2)}{\partial s_2},
\end{align*}
where $A_{ik} \in \R$ are constants to be determined. If we denote
the embedding map as $\Phi_{\text{inf}} \equiv (x_{ij})$ and
suppress the finite part, then $U_i$ action can be compactly
expressed as
\begin{equation}
\label{uop1}
(U_i f)(s_1,s_2) = e^{2 \pi i ( s_1 x_{3i} + s_2 x_{4i} )} f ( s_1
+x_{1i},s_2 +x_{2i}) .
\end{equation}
The condition (\ref{concr}) is satisfied if
\begin{equation} \label{embcon1}
x_{1i}x_{3i}+ x_{2i}x_{4i} =0 ,
\end{equation}
 and
\begin{equation} \label{Aik1}
A_{ik} =(\Phi_{\text{inf}}^{-1})_{ik} .
\end{equation}
Note that the above $U_i$ action (\ref{uop1}) would have an extra phase term
in the Manin's convention. However, the extra term,
 $x_{1i}x_{3i}+ x_{2i}x_{4i}$, has no contribution here
due to the condition (\ref{embcon1}).
Incorporating the effect of the finite part, we slightly change
the expression for the embedding map for the infinite part
(\ref{phii}) as follows:
\begin{align}\label{phi2}
\Phi_{\rm inf}=\begin{pmatrix}\theta_1 &0 &0&0\\
                0&0&\theta_2 &0\\
                0&1&0&0\\
                0&0&0&1\end{pmatrix} \equiv (x_{ij}).
\end{align}
Then, conditions (\ref{embcon1}) and (\ref{Aik1}) give
\[ ( A_{ik}) =\begin{pmatrix} \frac{1}{\theta_1 }&0&0&0\\
                0&0&1&0\\
               0&  \frac{1}{\theta_2 }&0&0\\
                0&0& 0&1 \end{pmatrix} . \]
Therefore the following operators specify a constant curvature
connection of right $ {\T}_\theta^4$-module:
\begin{align} \label{connection}
\nabla_1 & = - \frac{2 \pi i s_1 }{\theta_1 } ,
\notag \\
\nabla_2 & =  \frac{\partial}{\partial s_1} , \notag \\
\nabla_3 & = - \frac{2 \pi i s_2 }{\theta_2 } ,
\notag \\
\nabla_4 & =  \frac{\partial}{\partial s_2} .
\end{align}

The complexified connection space can be decomposed as a sum of a
holomorphic part and an antiholomorphic part. A complex structure
on the module $E$ can be introduced by choosing the
antiholomorphic subspace spanned by the following connection,
\begin{align}
\label{holconn}
\overline{\nabla}_1 & = \lambda_{11} \nabla_1 + \lambda_{12}
\nabla_2 + \lambda_{13} \nabla_3 + \lambda_{14} \nabla_4 , \notag \\
 \overline{\nabla}_2 & = \lambda_{21} \nabla_1 + \lambda_{22}
\nabla_2 + \lambda_{23} \nabla_3 + \lambda_{24} \nabla_4 , \notag
\end{align}
where $\lambda_{ij} \in \C .$ \ Choosing an appropriate basis such
that $(\lambda_{ij})$ becomes
\begin{equation*}
\begin{pmatrix} \lambda_{12} & \lambda_{14} \\
            \lambda_{22} & \lambda_{24} \end{pmatrix}^{-1}
            \begin{pmatrix}
           \lambda_{11}& \lambda_{12} &
             \lambda_{13} & \lambda_{14} \\
           \lambda_{21} & \lambda_{22} &
           \lambda_{23} & \lambda_{24}
            \end{pmatrix} = \begin{pmatrix} \tau_{11} & 1 & \tau_{12} & 0 \\
          \tau_{21} & 0 & \tau_{22} & 1 \end{pmatrix}
\end{equation*}
 the $(2 \times 2) $  matrix $\begin{pmatrix} \tau_{11} & \tau_{12}  \\
          \tau_{21}  & \tau_{22}  \end{pmatrix}, \ \tau_{ij} \in \C \ $ represents
          the complex structure of $ {\T}_\theta^4$-module.

Now we consider holomorphic vectors  in ${\T}_\theta^4$-module. A
vector $ f \in E $ is called holomorphic \cite{schwarz01} if it
satisfies
\begin{equation} \label{thetavector}
 \overline{\nabla}_i  f = 0 \ \ \text{for} \ \  i=1,2.
\end{equation}
 The above holomorphic
condition for $ f \in E $ now takes the form
\begin{align} \label{holcond1}
( \frac{2 \pi i \tau_{11} }{\theta_1 } s_1  +
 \frac{2 \pi i \tau_{12} }{\theta_2 } s_2
  ) f
 & =   \frac{\partial f}{\partial s_1} , \notag \\
 ( \frac{2 \pi i \tau_{21} }{\theta_1 } s_1  +
 \frac{2 \pi i \tau_{22} }{\theta_2 } s_2
   ) f
 & =  \frac{\partial f}{\partial s_2} .
\end{align}
%
In order for the two equations in (\ref{holcond1}) to be consistent
$\tau_{ij}$ should satisfy
\begin{equation} \label{holcondt4}
\frac{\tau_{12}}{ \theta_2 } = \frac{\tau_{21}} {\theta_1 }  .
\end{equation}
If $ {\rm Im} \Omega > 0$, Eq.(\ref{holcond1}) has a solution, the
so-called theta vector \cite{schwarz01,ds02} on noncommutative
${\T}^4$,
\begin{equation} \label{thetat4}
f (s_1 , s_2 ) = \exp [ \pi i S^t \Omega S ]
\end{equation}
where $ S=\begin{pmatrix} s_1 \\ s_2 \end{pmatrix}, s_i \in \R,
 \ i =1,2, $ and $ \Omega =
\begin{pmatrix} \frac{ \tau_{11} }{\theta_1 } &
\frac{ \tau_{12} }{\theta_2 } \\
\frac{ \tau_{21} }{\theta_1 } & \frac{ \tau_{22}
}{\theta_2 }
\end{pmatrix} $.
\\


\subsubsection*{3.1.2 Embedding into lattice}
%




 For the embedding of the type $ {\R} \times {\Z}^2 $ the
  relation (\ref{concr}) is satisfied if we let
\begin{align}
(\nabla_i f)(s,n_1,n_2)&= - 2\pi iB_{i1}s f(s,n_1,n_2) - 2\pi
iB_{i2}n_1 f(s,n_1,n_2) \notag \\
& ~~ - 2\pi iB_{i3}n_2 f(s,n_1,n_2) + B_{i4}\frac{\partial
f(s,n_1,n_2)}{\partial s}, ~~ \text{for} ~~ i=1,..,4,
\label{bmatrix}
\end{align}
where $B_{ik} \in \R$ are constants satisfying the following
condition:
\begin{equation} \label{amatrix}
B_{i1}x_{1j}+ B_{i2}x_{2j} + B_{i3}x_{3j} +B_{i4}x_{4j} =
\delta_{ij} , ~~~ i,j =1,..,4,
\end{equation}
while $x_{ij}$'s in (\ref{embqnz}) should satisfy the following condition,
\begin{equation} \label{embcon}
x_{1j}x_{4j}+ x_{2j}x_{5j} + x_{3j}x_{6j} =0, ~~~ j= 1,..,4 .
\end{equation}



\noindent The embedding map  (\ref{embqnz}) satisfies the
condition (\ref{embcon}), and the condition (\ref{amatrix}) gives
\[ ( B_{ik}) =\begin{pmatrix} \frac{1}{\theta_1 }&0&0&0\\
                0&0&0&1\\
               0& b_{11}  & b_{12}&0\\
                0& b_{21} & b_{22} &0 \end{pmatrix} , \]
where
\begin{align} \label{invbm}
 \begin{pmatrix}
                b_{11}  & b_{12}\\
                 b_{21} & b_{22}  \end{pmatrix} =
                 { \begin{pmatrix}
                m_{11}  & m_{12}\\
                 m_{21} & m_{22}  \end{pmatrix} }^{-1} .
\end{align}
Therefore the following operators specify a constant curvature
connection of right $ {\T}_\theta^4$-module $E$:
\begin{align} \label{connection}
\nabla_1 & = - \frac{2 \pi i s }{\theta_1 } ,
\notag \\
\nabla_2 & =  \frac{\partial}{\partial s} , \notag \\
\nabla_3 & = - 2 \pi i (b_{11} n_1 + b_{12} n_2)  ,
\notag \\
\nabla_4 & = - 2 \pi i (b_{21} n_1 + b_{22} n_2)  .
\end{align}

A complex structure on the module $E$ might be introduced
in the same manner as in the
previous case:
\begin{align} \label{holconn}
\overline{\nabla}_1 & = \lambda_{11} \nabla_1 + \lambda_{12}
\nabla_2 + \lambda_{13} \nabla_3 + \lambda_{14} \nabla_4 , \notag \\
 \overline{\nabla}_2 & = \lambda_{21} \nabla_1 + \lambda_{22}
\nabla_2 + \lambda_{23} \nabla_3 + \lambda_{24} \nabla_4 , \notag
\end{align}
where $\lambda_{ij} \in \C .$ \ And choosing an appropriate basis
$(\lambda_{ij})$ can be expressed as
\begin{equation*}
\begin{pmatrix} \lambda_{13} & \lambda_{14} \\
            \lambda_{23} & \lambda_{24} \end{pmatrix}^{-1}
            \begin{pmatrix}
           \lambda_{11}& \lambda_{12} &
             \lambda_{13} & \lambda_{14} \\
           \lambda_{21} & \lambda_{22} &
           \lambda_{23} & \lambda_{24}
            \end{pmatrix} =
\begin{pmatrix} \tau_{11}  & \tau_{12}& 1 & 0 \\
          \tau_{21} & \tau_{22} & 0 & 1 \end{pmatrix} ,
\end{equation*}
with the $(2 \times 2) $  matrix $\begin{pmatrix} \tau_{11} & \tau_{12}  \\
          \tau_{21}  & \tau_{22}  \end{pmatrix}, \ \tau_{ij} \in \C \ $ representing
          the complex structure of $ {\T}_\theta^4$-module.

To be a holomorphic vector  in ${\T}_\theta^4$-module,
 $ f \in E $ now takes the form
\begin{align} \label{holcond2}
 & ( \frac{2 \pi i \tau_{11} }{\theta_1 } s  + 2 \pi i (b_{11} n_1 + b_{12} n_2)
  ) f
  =  \tau_{12} \frac{\partial f}{\partial s} , \notag \\
 & ( \frac{2 \pi i \tau_{21} }{\theta_1 } s +  2 \pi i (b_{21} n_1 + b_{22} n_2) ) f
  =  \tau_{22} \frac{\partial f}{\partial s}  .
\end{align}
%
In order for the two equations in (\ref{holcond2}) to be
consistent, $\tau_{ij},  b_{ij}$ should satisfy
\begin{align*}
\frac{\tau_{11}} {\tau_{12}} = \frac{\tau_{21}} {\tau_{22}}, ~~
 b_{12} = \frac{\tau_{12}} {\tau_{22}} b_{22}, ~ b_{21} = \frac{\tau_{22}} {\tau_{12}} b_{11}  .
\end{align*}
However, the above result yields
\[ \det \begin{pmatrix}
                b_{11}  & b_{12}\\
                 b_{21} & b_{22}  \end{pmatrix} = 0 ,  \]
which is contradictory to the assumption that $ (b_{ij} )$ is the
inverse matrix of $ (m_{ij})$, the relation (\ref{invbm}).

The above shows that one cannot have a holomorphic vector over
totally complexified ${\T}_\theta^4 $ in the embedding of $ {\R}
\times {\Z}^2 $. This can be remedied by giving a complex
structure only over the continuous part of the embedding space,
i.e., by giving a complex structure to the connection components
over $ {\R} \times {\R}^* $.
Now, we implement this as follows.
\begin{align}
\label{holconn2}
\overline{\nabla}_1 & = \tau \nabla_1 + \nabla_2 , \notag \\
 \overline{\nabla}_2 & =  \nabla_3 , \notag \\
\overline{\nabla}_3 & = \nabla_4 ,
\end{align}
where $\tau \in {\C} $ is a complex structure constant over $ {\R}
\times {\R}^* $. Then, the holomorphic vectors over this complex
structure satisfy
\begin{align}
\label{holtheta2}
\overline{\nabla}_1  f (s, n_1, n_2 )  = 0 ,
\end{align}
which is
\[
      - \frac{2 \pi i \tau }{\theta_1 } s f  +   \frac{\partial f}{\partial s} = 0 .
\]
Since $f$ belongs to $ {\mathcal S}({\mathbb R}) \otimes {\mathcal
S} ( {\Z}^2 ) $, $f(s, n_1, n_2)$ satisfying (\ref{holtheta2}) can
be given by
\begin{equation}
\label{theta2}
f (s, n_1, n_2  ) = \exp (  \frac{ \pi i \tau }{\theta_1 } s^2  ) g (n_1, n_2)
\end{equation}
where $g (n_1, n_2) \in {\mathcal S} ( {\Z}^2 )$ is a Schwartz
function. For the function $g(n_1, n_2)$, we will use a simple
Schwartz function such that $ f (s, n_1, n_2 )$ can be expressed
as
\begin{equation}
\label{thevec2}
 f (s, n_1, n_2  ) = \exp [  \pi i  \frac{\tau
}{\theta_1 } s^2 - \pi \frac{1}{\theta_2} (n_1^2 + n_2^2)  ] ,
\end{equation}
where ${\rm Im} \tau >0,$ and $\theta_1 = \theta_{12}>0 , ~
\theta_2 = \theta_{34} >0$ are
     given in  (\ref{Ucomm}).



\subsection*{3.2 Quantum theta functions}\label{qthftn}


 Before considering quantum theta function,
 we first review the algebra valued inner product on a bimodule
after Rieffel \cite{rief88}.
Let $M$ be any locally compact Abelian group, and $\widehat{M}$ be
its dual group, and let ${\cal G} \equiv M \times \widehat{M} $.
Let $\pi$ be a representation of ${\cal G}$ on $L^2(M)$ such that
\begin{align}
\pi_x \pi_y = \alpha (x,y) \pi_{x+y} =\alpha (x,y)
\overline{\alpha}(y,x) \pi_y \pi_x ~~~ {\rm for}~~ x,y \in {\cal
G} \label{ccl}
\end{align}
where $\alpha$ is a map $ \alpha : ~ {\cal G} \times {\cal G}
\rightarrow {\C}^* $ satisfying
\[ \alpha(x,y)
=\alpha(y,x)^{-1} , ~~~ \alpha(x_1 + x_2 , y) = \alpha(x_1 , y)
\alpha (x_2 , y) ,  \] and $\overline{\alpha}$ denotes the complex
conjugation of $\alpha$.
%
 Let $D$ be a
discrete subgroup of $\cal{G}$. We define $\mathcal{S}(D)$ as the
space of Schwartz functions on $D$.
 For $\Psi \in \mathcal{S}(D)$, it can be expressed as $\Psi = \sum_{w \in D} \Psi(w) e_{D,
 \alpha}(w)$ where $e_{D, \alpha}(w)$ is a delta function with
 support at $w$ and obeys the following relation.
\begin{equation}
e_{D, \alpha} (w_1) e_{D, \alpha} (w_2) = \alpha(w_1,w_2) e_{D,
\alpha} (w_1 +w_2) \label{ccld}
\end{equation}
%


 For Schwartz functions $f,g \in \mathcal{S}(M)$, the algebra ($\mathcal{S}(D)$) valued
inner product is defined as
\begin{align}
{}_D <f,g> \equiv \sum_{w\in D} {}_D<f,g>(w) ~ e_{D, \alpha}(w) ~
\label{aip}
\end{align}
where
\begin{align}
{}_D<f,g>(w) = <f, \pi_w g> . \nonumber
\end{align}
Here, the scalar product of the type $<f,p>$ above with $f,p \in
L^2 (M)$ denotes the following.
\begin{align}
<f,p> = \int f(x_1) \overline{p(x_1)} d \mu_{x_1}  ~~~{\rm for} ~~
x=(x_1,x_2) \in M \times \widehat{M} , \label{sp}
\end{align}
where $\mu_{x_1}$ represents the Haar measure on $M$ and
$\overline{p(x_1)}$ denotes the complex conjugation of $p(x_1)$.
 The $\mathcal{S}(D)$-valued inner product can be represented as
\begin{align}
{}_D <f,g> =\sum_{w\in D} <f, \pi_w g> ~ e_{D, \alpha}(w) ~.
\label{aipr}
\end{align}
%
 For $\Psi \in \mathcal{S}(D) $ and $f \in
\mathcal{S}(M)$, then $~ \pi (\Psi) f \in \mathcal{S}(M)$ can be
written as \cite{rief88}
\begin{align}
(\pi(\Psi)f)(m) & = \sum_{w \in D} \Psi (w) (\pi_w f) (m)
\end{align}
where $m\in M, ~ w \in D \subset M \times \widehat{M}$.

\subsubsection*{3.2.1 Embedding into vector space}

Now, we consider Manin's quantum theta function $\Theta_D$
\cite{manin1-tr,manin2,manin3} for the embedding into vector
space. In \cite{manin3}, the quantum theta function was defined
via algebra-valued inner product up to a constant factor
\cite{ek1},
\begin{align}
{}_D<f , f > &  \sim  \Theta_D , \label{qtheta-def}
\end{align}
where $f$ used in the Manin's construction \cite{manin3} was a
simple Gaussian theta vector
\begin{align}
f = e^{\pi i x_1^t T x_1}, ~~ x_1 \in M. \label{tv-gauss}
\end{align}
Here $T$ is a complex structure given by a complex  skew symmetric
matrix.
%
%
With a given complex structure $T$, a complex variable
$\underline{x} \in {\C}^n$  can be introduced via
\begin{align}
\underline{x} \equiv T x_1 +x_2
\end{align}
where $x=(x_1, x_2) \in M \times \widehat{M}$.

 Based on the defining
concept for quantum theta function (\ref{qtheta-def}), one can
define the quantum theta function $\Theta_D$ in the noncommutative
${\T}^4$ case as
\begin{align}
{}_D<f , f > & = \frac{1}{\sqrt{2^2 \det ({\rm Im} ~ T )}}
\Theta_D \label{qtfM}
\end{align}
for $f$ given by (\ref{tv-gauss}) and $T$ given by $\Omega$ that
appeared in (\ref{thetat4}).
According to (\ref{aip}), the $\mathcal{S}(D)$-valued inner
product  (\ref{qtfM}) can be written as
\begin{align}
{}_D<f , f >  =\sum _{h \in D} <f , \pi_h f > e_{D, \alpha} (h) .
\label{sdip}
\end{align}
  In \cite{manin3}, Manin  showed that the quantum theta
function defined in (\ref{qtfM}) is given by
\begin{align}
 \Theta_D & = \sum _{h \in D}  e^{- \frac{\pi}{2} H(\underline{h},\underline{h}) }
  e_{D, \alpha} (h) ,
\label{TDM}
\end{align}
where
\[
H( \underline{g}, \underline{h} ) \equiv \underline{g}^t ( {\rm
Im} T)^{-1} \underline{h}^*
\]
with $ \underline{h}^* = \overline{T} h_1 + h_2 $ denoting the
complex conjugate of $\underline{h}$. At the same time, it also
satisfies a quantum version of the translation action for the
classical theta functions \cite{manin1-tr}:
\begin{equation}
{}^\forall g \in D, ~~ C_g ~ e_{D, \alpha} (g) ~ x_g^* ( \Theta_D)
= \Theta_D
\label{TDfnr}
\end{equation}
where $C_g$ is defined by
\[ C_g = e^{- \frac{\pi}{2} H(\underline{g},\underline{g})} \]
and the action of $x_g^*$, 'quantum translation', is given by
\begin{align}
x_g^* (e_{D, \alpha} (h)) = e^{- \pi
H(\underline{g},\underline{h})} e_{D, \alpha} (h).
\label{xtrans}
\end{align}
In \cite{manin1-tr}, Manin has also required that the factor $
C_g, ~ g \in D $ appearing in the quantum translation $ x_g^* $
has to satisfy the following relation under a combination of
quantum translations for consistency.
\begin{align}
\frac{C_{g+ h}}{C_g C_h} =  {\cal T}_g(h) \alpha(g,h).
\label{xtr-cond}
\end{align}
Here $\alpha(g,h)$ is the cocycle appearing in (\ref{ccld}), and
${\cal T}_g(h)$ is a generalized expression of the factor that
appears by quantum translation:
\begin{align}
x_g^* (e_{D, \alpha} (h)) \equiv  {\cal T}_g(h) e_{D, \alpha} (h).
\label{xtrans-gen}
\end{align}
 The proof of the functional relation (\ref{TDfnr}) in this embedding case
 with quantum translation (\ref{xtrans})
  was shown in \cite{manin3}, in which the complex structure
$T$ is given by  $\Omega$  in (\ref{thetat4}).

\subsubsection*{3.2.2 Embedding into lattice}

 We now turn to the second embedding case of nonzero $q$, where we do not have
  holomorphic vectors, the so-called theta
 vectors, once we assign a complex structure over the whole
${\T}^4$.
A way-out from this difficulty turned out to be introducing a
complex structure partially, i.e., only over the continuous
subspace of the embedding space. As a result of this we got the
function $f(s,n_1,n_2)$ (\ref{thevec2}) as an element of the
module relevant to the  nonzero $q$ embedding.

With the function $f(s,n_1,n_2)$, we now evaluate the quantum
theta function, and see whether it satisfies the functional
relation for 'quantum translation'.
We first define the quantum theta function  {\it $\grave{a}$ la}
(\ref{qtfM}) for ${\T}^4$ in the $q=2$ case:
\begin{align}
\frac{1}{\sqrt{2  {\rm Im} ~ T }} \hat{\Theta}_{D}  =  {}_D<f ,
f > , \label{qtfM2}
\end{align}
where $T$ is  a 'complex structure' over the continuous part of
the embedding space to be specified below.
We then show that the above-defined quantum theta function
satisfies a functional relation {\it $\grave{a}$ la} (\ref{TDfnr}) with
modified quantum translation :
\begin{equation}
{}^\forall g \in D, ~~ \hat{C}_g ~ e_{D, \alpha} (g) ~ \hat{x}_g^*
( \hat{\Theta}_{D}) = \hat{\Theta}_D
\label{TDfnr2}
\end{equation}
where $\hat{C}_g, ~ \hat{x}_g^* $ are to be defined below.

To evaluate the quantum theta function (\ref{qtfM2}), we calculate
the scalar product inside the summation in (\ref{sdip}) first.
 For that we first write the action of the operator $\pi_h $ on $f$:
\begin{equation}
\pi_h f(s,n_1,n_2)   =  e^{ 2 \pi i ({w_h}_2 s + t_1 n_1 + t_2
n_2) + \pi i ( {w_h}_1 {w_h}_2  + m_1 t_1  + m_2 t_2 )  } f
(s+{w_h}_1 ,n_1 +m_1,n_2+m_2) , \label{actionforpi}
\end{equation}
where $h \in D$ is given by
\[ h = ( {w_h}_1, {w_h}_2, m_1, m_2, t_1, t_2 ) \in {\R} \times
{\R}^* \times {\Z} \times {\Z} \times {\T} \times {\T} . \]
Then,
\begin{eqnarray}
<f , \pi_h f > & = & \sum_{n_1,n_2 \in {\Z}} \int_{{\R}} ds ~ e^{
\pi [ i \frac{\tau}{\theta_1} s^2 - \frac{1}{\theta_2} (n_1^2
+n_2^2)]}
  e^{- 2 \pi i ({w_h}_2 s + t_1 n_1 + t_2 n_2) - \pi i ( {w_h}_1
{w_h}_2  + m_1 t_1  + m_2 t_2 )  }
\nonumber \\
&  & ~~~ \hspace*{1cm} \times e^{  \pi [ - i
\frac{\bar{\tau}}{\theta_1} (s+ {w_h}_1)^2 - \frac{1}{\theta_2}
[(n_1 +m_1)^2 +(n_2 +m_2)^2]]}
\nonumber \\
& = &  \int_{{\R}} d s ~ e^{ -2\pi [ \frac{{\rm Im}\tau}{\theta_1}
s^2 + i \frac{\bar{\tau} }{\theta_1} {w_h}_1 s + i {w_h}_2 s] - i
\pi [\frac{\bar{ \tau}}{\theta_1} ({w_h}_1)^2 + {w_h}_1 {w_h}_2 ]}
\nonumber \\
&  &  \times  e^{- \frac{\pi}{\theta_2}(m_1^2 + m_2^2)
  - \pi i ( m_1 t_1 + m_2 t_2 ) }  \sum_{n_1,n_2 \in {\Z}}
 e^{- \frac{2 \pi}{\theta_2} (n_1^2 + n_2^2) +  2 \pi i [ n_1 ( -t_1 +  \frac{i m_1}{\theta_2})
     +  n_2 ( -t_2 +  \frac{i m_2}{\theta_2})]}
 \nonumber \\
& = &  b_{t_1, m_1} b_{t_2,m_2}  \int_{{\R}} d s ~ e^{ -2\pi
[\frac{{\rm Im}\tau}{\theta_1} s^2 + i \frac{\bar{\tau}
}{\theta_1} {w_h}_1 s + i {w_h}_2 s] - i \pi [\frac{\bar{
\tau}}{\theta_1} ({w_h}_1)^2 + {w_h}_1 {w_h}_2 ]} , \label{scprod}
\end{eqnarray}
where
\begin{equation}
 b_{t_j, m_j} =  e^{- \frac{\pi}{\theta_2} m_j^2 - \pi i m_j t_j}
                ~ \theta (\tau = \frac{2 i}{\theta_2}, ~ z= -t_j + \frac{im_j}{\theta_2}),
  ~~ j=1,2 .
\label{c-latt}
\end{equation}
Here, $\theta (\tau , z)$ is the classical theta function defined
by
\[
\theta (\tau ,  z) = \sum_{n \in {\Z}} e^{ \pi i  \tau n^2 +  2
\pi i n z }, ~ ~ {\rm for} ~ ~ \tau , z \in {\C} .
\]
In order to facilitate the integration part, we denote the
integrand as
\[ e^{-\pi [q(s)+ l_{w_h} (s)+ \widetilde{C}_{w_h}]}  \]
with
\begin{align*}
q(s) & = 2 ({\rm Im} T) ~s^2 , \\
l_{w_h}(s) & = 2 i  (T^* {w_h}_1  + {w_h}_2 ) s ,   \\
\widetilde{C}_{w_h} & =i {w_h}_1 ( T^* {w_h}_1 + {w_h}_2 ) ,
\end{align*}
where
\[ T = \frac{\tau}{\theta_1} ~ .  \]
 Using the relation
 \[  q(s + \lambda_{w_h}) -q(\lambda_{w_h}) = q(s) + l_{w_h} (s)  \]
with
 \[ \lambda_{w_h} \equiv \frac{i}{2} ( {\rm Im} T)^{-1} \underline{w_h}^* , \]
the integration  becomes
\begin{eqnarray*}
  \int_{\R} d s ~ e^{- \pi (q(s) + l_{w_h}(s) + \widetilde{C}_{w_h} )}  =
  e^{- \pi ( \widetilde{C}_{w_h} - q(\lambda_{w_h}))}
 \int_{\R} d s ~  e^{- \pi q( s  + \lambda_{w_h})}
 = \frac{1}{\sqrt{2 {\rm Im} T }} e^{- \pi (\widetilde{C}_{w_h} -q(\lambda_{w_h}))} .
\end{eqnarray*}
With a straightforward calculation one can check that
\[
\widetilde{C}_{w_h} -q(\lambda_{w_h}) = \frac{1}{2} H(\underline{w_h},
\underline{w_h}) .
\]
Thus the quantum theta function $\hat{\Theta}_D$ is given by
\begin{align}
 \hat{\Theta}_D & = \sum _{h \in D} \widetilde{b}_h ~ e^{- \frac{\pi}{2} H(\underline{w_h},\underline{w_h}) }
  e_{D, \alpha} (h) ,
\label{TDM2}
\end{align}
where
\begin{equation}
\widetilde{b}_h = \prod_{j=1}^2 b_{t_j, m_j}
\label{lat-factor}
\end{equation}
with $b_{t_j, m_j}$ given in (\ref{c-latt}).
To be consistently maintaining the symmetry property of classical
theta function under lattice translation, the above given quantum
theta function should satisfy the functional relation under
'quantum translation' (\ref{TDfnr2}),
\begin{equation*}
{}^\forall g \in D, ~~ \hat{C}_g ~ e_{D, \alpha} (g) ~ \hat{x}_g^*
( \hat{\Theta}_{D})
 = \hat{\Theta}_{D} ,
\end{equation*}
and the consistency condition (\ref{xtr-cond}) for $\hat{C}_g$.
The above relation is satisfied if we assign
\begin{equation}
\hat{C}_g = \widetilde{b}_g ~ e^{- \frac{\pi}{2}
H(\underline{w_g},\underline{w_g})} ,
\label{trc-lat}
\end{equation}
and $ \hat{x}_g^* $ is defined by
\begin{equation}
\hat{x}_g^* (e_{D, \alpha} (h)) = \hat{{\cal T}}_g(h) e_{D,
\alpha} (h) \label{tr-latt}
\end{equation}
with
\begin{align}
\hat{{\cal T}}_g(h) = \frac{\hat{C}_{g+ h}}{\hat{C}_g \hat{C}_h
\alpha(g,h) } .
\label{xtr2-cond}
\end{align}
Here we note that the quantum translations are not additive in
this case:
\begin{align}
\hat{x}_{g_1}^* \cdot \hat{x}_{g_2}^* (e_{D, \alpha} (h)) \neq
\hat{x}_{g_1 + g_2}^* (e_{D, \alpha} (h)) .
\label{qtr-latt}
\end{align}
On the other hand, the quantum translations in the Manin's case
($x_g^*$), (\ref{xtrans}), are additive:
\begin{align}
x_{g_1}^* \cdot x_{g_2}^* (e_{D, \alpha} (h)) = x_{g_1 + g_2}^*
(e_{D, \alpha} (h)) .
\label{qtr-vectsp}
\end{align}
Now, it is easy to show the relation (\ref{TDfnr2}):
\begin{eqnarray*}
 \hat{C}_g ~ e_{D, \alpha} (g) ~ \hat{x}_g^* ( \hat{\Theta}_{D})
& = & \hat{C}_g ~ e_{D, \alpha} (g) ~ \hat{x}_g^* (\sum _{h \in D}
\widetilde{b}_h ~ e^{- \frac{\pi}{2}
H(\underline{w_h},\underline{w_h}) }
  e_{D, \alpha} (h)) \\
& = & \hat{C}_g ~ e_{D, \alpha} (g) ~ \hat{x}_g^* (\sum _{h \in D}
\hat{C}_h e_{D, \alpha} (h))\\
& = & \sum _{h \in D} \hat{C}_g \hat{C}_h e_{D, \alpha} (g)
\hat{{\cal T}}_g(h) e_{D, \alpha} (h)\\
 & = &  \sum _{h \in D} \hat{C}_{g+ h} e_{D, \alpha} (g+h)   = \hat{\Theta}_{D} .
\end{eqnarray*}
where we used the relation (\ref{trc-lat}) in the second step, and
the relation (\ref{xtr2-cond}) together with the cocycle condition (\ref{ccld})
in the last step.
\\

%
%




\section*{4. Conclusion }

In this paper, we  study the theta vector and the corresponding
quantum theta function in the embedding into lattice for the
noncommutative 4-torus.

While the theta vector exists in the embedding into the vector
space case ($ {\R}^p $ type), it does not exist in the embedding
into the lattice case (${\Z}^q $ type). And thus holomorphic theta
vectors only exist for the vector space part in the case of mixed
embedding ($ {\R}^p \times {\Z}^q $ type). In general, the modules
from embeddings including the lattice part are not fully
holomorphic. Manin constructed the quantum theta functions only
with holomorphic modules. Therefore, it is natural to ask whether
one can construct the quantum theta function satisfying the
Manin's requirement with the partially holomorphic modules in the
mixed embedding case.

It turns out that these non-holomorphic modules also satisfy the
requirement of the quantum theta function of Manin. We show this
explicitly for the noncommutative 4-torus case with embedding into
$ {\R} \times {\Z}^2 $.
However, we note a feature that is different among the two quantum
theta functions.  In our quantum theta function constructed with a
partially-holomorphic module, two consecutive 'quantum
translations' are not additive, while those in the Manin's are
additive. This happens due to the consistency condition between
quantum translation and cocycle, condition (\ref{xtr-cond}).
The same holds for the quantum theta functions constructed with
modules from embeddings into lattice (${\Z}^q$) part only.
This is due to the structure of the quantum theta function shown
in (\ref{TDM2}) and that of the coefficient of the quantum
translation, (\ref{trc-lat}). Both of them consist of a direct
product of the contributions from the two parts, one from the
embedding into vector space and the other from the embedding into
lattice.

In conclusion, we show explicitly that the quantum theta function
that Manin defined can be constructed with any choice of the
following embeddings, (1) into vector space times lattice, (2)
into vector space, (3) into lattice, for the noncommutative
4-torus.
We expect that this will hold for higher dimensional
noncommutative tori.
\\

\pagebreak

\noindent
{\Large \bf Acknowledgments}

\vspace{5mm} \noindent The authors thank KIAS for hospitality
during the time that this work was done. This work was supported
by Korean Council for University Education, grant funded by Korean
Government(MOEHRD) for 2006 Domestic Faculty Exchange (E C-Y), and
by KOSEF Research Grant No R01-2006-000-10638-0 (H K).
\\





\begin{thebibliography}{88}

\bibitem{schwarz01} A. Schwarz, {\it Theta-functions on noncommutative
tori},  Lett. Math. Phys. 58,  81 (2001).

\bibitem{manin1} Y. Manin, {\it Quantized theta-functions} in:
Common trends in mathematics and quantum field theories (Kyoto,
1990), Progress of Theor. Phys. Suppl. 102, 219 (1990).

\bibitem{manin1-tr} Y. Manin,
 {\it Theta functions, quantum tori and Heisenberg groups}, math.AG/0011197.

\bibitem{manin2} Y. Manin, {\it Real multiplication and
noncommutative geometry}, math.AG/0202109.

\bibitem{manin3} Y. Manin, {\it Functional equations for quantum
theta functions}, math.QA/0307393.

\bibitem{rief88} M. Rieffel, {\it Projective modules over higher-dimensional non-commutative tori},
 Can. J. Math. Vol. XL, 257 (1988).

\bibitem{ds02} M. Dieng and A. Schwarz, {\it Differential and complex
geometry of two-dimensional noncommutative tori}, QA/0203160.

\bibitem{ek05} Ee C.-Y. and H. Kim, {\it Symmetry of quantum torus
with crossed product algebra}, J. Math. Phys. 47, 073058 (2006).

\bibitem{kl03} H. Kim and C.-Y. Lee, {\it Theta functions on
 noncommutative $T^4$},  J. Math. Phys. 45, 461 (2004).

\bibitem{rs98}M. Rieffel and A. Schwarz, {\it Morita equivalence of multidimensional noncommutative tori},
Int. J. Math. 10, 289 (1999).




\bibitem{ek1} Ee C.-Y. and H. Kim, {\it Theta vectors and quantum theta functions}, J. Phys. A 38, 4255 (2005).



\end{thebibliography}
\end{document}